\documentclass[11pt]{article}
\pdfoutput=1
\usepackage{graphicx}

\begin{document}

\title{Determination of the helpfulness of physics exam study methods}

\author{Rahul Jilakara and David P. Waters\\Department of Basic Sciences, University of Health Sciences and Pharmacy\\St. Louis, Missouri 63110, USA.}



\begin{abstract}

Studying for physics exams can be difficult and stressful, especially during a student’s introductory year in physics. For students who do not plan to major in physics, the desire to do well is based less on understanding concepts and more on achieving a better grade. For this reason, students want to study as efficiently as possible by using the most optimal study methods. We have taken surveys over the past three years to determine how students study for exams and compared that to their exam grades. We found that students who studied using methods that they rated as more helpful did better on the exams. By utilizing the study results, we are able to present our current and future students with study methods that have been rated as being more helpful, and give them advice on ways to optimize their study time for exams.

\end{abstract}

\maketitle

\section{BACKGROUND}

As we prepare this study, we recognize that the information gathered was based on self-report and Likert-type instruments. Self-report means that a student provides their own ratings instead of using an independent determination, and these have both advantages and disadvantages \cite{Schellings11,Hout09, Veenman11}. The advantages of self-reports are that we do not interrupt or influence someone during their process, and it allows us to collect a lot of data relatively easily. Unfortunately, students may not be able to accurately recall everything, or may be biased in their answers \cite{Veenman05}. Althubaiti et al. have looked at the information bias in health research using self-reporting, and have found that recognizing these biases can help to lessen the effects \cite{Althubaiti16}. 

Study habits are important for college students, and not all students know how to study in an optimal manner \cite{Kumar15}. For example, research has been conducted on how best to study for exams in medical school \cite{Augustin14}, which looked at the testing effect, active recall, and spaced repetition, which are all examples of study skills. In our study, we will be looking at study methods that come from preparing for physics examinations during introductory algebra-based physics courses. We define study methods as the materials that students use to study for exams.

It would be ideal if there were a set of universally helpful study methods that we could provide to our students. Unfortunately, research has shown that different students may use this information more or less successfully. For example, high-achieving students have been found to  better prepare for exams by using study methods that are more helpful \cite{ Kitsantas02, Pintrich02, Sundre04, VanZile99} while low-achieving students used the order of materials to determine what they would need to study \cite{Holschuh00}. An interesting finding is that some low-achieving students would recommend more helpful study strategies - a combination of study skills and study methods - to others, but then would use less effective study strategies for themselves \cite{Holschuh00}. These students often would not use more effective strategies because of a lack of time or motivation \cite{Barnett00}. These findings show why a student would continue to use a study method that they continuously rated as less helpful. It is also possible that a student’s ability to recognize whether something was helpful could be biased \cite{Etten97}.

Previous work on study strategies for exam preparation has found that providing information about how to study can be more useful than providing the exact questions on the exam \cite{Etten97}. The authors found that distributing efforts across many study methods can help in preparing for an exam. Also, they found that knowing the format of the exam can be helpful for students. For this reason, our physics courses provide last year’s exam to students, in order to both show the format of the exam and give students material to prepare for the exam. Fakcharoenphol et al. find that studying using old exam practice problems can help, although learning may be shallow \cite{Fak11}. Since our exams come right after studying, we do not know if the learning that brought about higher exam scores was salient. Previous research also determined that studying on their own using worked-out solutions – old exams and class slides (problems) – and targeted exercises such as the review problems allowed students to perform better than they did when using traditional study methods, such as rereading the textbook \cite{Fak14}. Finally, completing web-based physics homework problems can lead to higher overall exam performance \cite{Dufresne02}, so using these homework problems to study may also help students improve their exam scores.

Although many study methods can improve a student’s exam scores, some study methods have been found to be less helpful. Although students believed that reading the textbook in order to study for an exam improved their understanding, doing so had no effect on their exam performance because it gave them a false sense of understanding \cite{Linder10}. This finding is interesting because many studies that look at exam preparation only have textbook readings as their study method \cite{Pressley97}. Pressley et al. determines strategies to use textbook reading successfully, which may explain why some of our students label textbook reading as helpful.

It has been found that students must adapt their study method to the type of task that they are expected to complete \cite{Broekkamp07}, so a study method that results in better exam grades for most students may not be as helpful in another environment. It may be difficult to make a list of helpful study methods from our physics courses and implement them in other environments. One would expect that some study methods would be more helpful than others. Some students are going to learn better from certain study methods while others will not learn as well using the same study methods \cite{Pressley97}. This may explain why study methods in our courses work better for some students than for others. Assuming that our students’ self-reporting is accurate, it may be that students use the study methods differently or that they begin the preparation with different levels of understanding, and thus will need to use different methods.

\section{PROCEDURE}
\subsection{Population}
The survey was given to undergraduate students, mostly sophomores and juniors, at the University of Health Science and Pharmacy in St. Louis. The courses in the fall and spring were algebra-based introductory physics courses. These courses were required for health science majors who were most often going to enter the pharmacy program or take a health science professional exam such as the MCAT. The fall and spring courses each had a typical enrollment of 60-80 students.

\subsection{\label{sec:level2}Data Collection}
After each exam, students were asked to fill out a survey. These surveys were administered using the course’s LMS (Moodle). The surveys were available to all students, but participation was voluntary.

The survey questions are listed below:

\begin{itemize}
   \item \textbf{Q1}: Approximately how many hours did you study for the exam?
   \item \textbf{Q2}: What percentage of time did you use the following materials to study for the exam? (10\%, 20\%, etc. and N/A)
   \item \textbf{Q3}: How helpful were these material types for the exam? (Scale: 1 = least helpful and 5 = most helpful)
\end{itemize}

Any student who did not use a certain study method answered N/A on Q2 and Q3 for that study method. Some students did not choose N/A for Q3 for a study method that they did not use, so we changed that Q3 value to an N/A for them. 

A list of the study methods available along with an explanation of each was given:

\begin{itemize}
    \item \textbf{Rereading Textbook} - The course uses OpenStax College Physics Textbook \cite{Textbook}. Sections of the textbook are given as a smaller PDF before each class. The assumption is that students read the textbook before class, so this method would involve reading the textbook again before the exam.
    \item \textbf{Class Slides (Conceptual)} - During each class, students are asked multiple choice questions and use a clicker to input the answer. These questions are conceptual in nature, meaning that there usually is not any math needed to answer them. They are similar to the multiple choice questions on the exam.
    \item \textbf{Class Slides (Problems)} - Each week, after the topics have been introduced and discussed, students are given a day to work only on mathematical problems in class. These problems are similar to the mathematical problems that students will see on the exam. Worked-out solutions are provided.
    \item \textbf{Review Questions} - Review questions are posted for the students to work on. These involve multiple choice questions that may be either conceptual or mathematical in nature. These are all new questions that have not been seen before. The answers are provided, but the solutions are not. For AY 2017-2018, the questions were on our LMS, but the number of questions was limited. For AY 2018-2019, these questions were posted externally with more questions added. In AY 2019-2020, these questions were posted on the course’s LMS page with many more questions available.
    \item \textbf{Old Exam} - The previous year’s exam is provided as a study method, both to give insight about the format of the exam as well as to provide more practice opportunities. The answers are provided.
    \item \textbf{Homework Problems} - Homework assignments involve solving mathematical problems similar to those seen in class and those on the exam.
    \item \textbf{Reading quizzes} - Before each class, students are asked a series of questions to see if they have read the textbook and/or watched the video. These are simple questions to check for a basic understanding of the pre-class material.
\end{itemize}

\subsection{Course Exams}
The course exams were made up of conceptual (multiple-choice) questions and mathematical problems. These were paper exams on which students would write out their work. For the multiple choice questions, students could write out their work for partial credit. The mathematical problems also allowed for partial credit, with the bulk of the points awarded for showing how to solve the problem. All exam scores included in the data are after partial credit was given. Exams usually had about 10 MC questions worth 3-4 points each and 3-4 problems with individual parts that made up more than half of the grade. For our analysis, we include the overall grade for the exam as well as scores for the conceptual section and scores for the problems section. This separation allowed us to compare study methods that focused on concepts versus study methods that required mathematical solutions.

\section{Study Method Usage}
\subsection{Data Analysis}

The collected data from the student-filled surveys was divided and analyzed according to the semester it was collected. For each semester, we determined the average amount of time that a student studied for each particular exam from Q1. Likewise, using the data from Q2, we averaged the percent of time that each student spent utilizing each of the seven study habits per exam over the course of the semester.

Next, we took the percent of time spent on each study method from each semester and averaged it all together to produce the percent of time spent on each study method for the overall collected data. The percent of time spent per each study method was then multiplied by the average amount of time spent studying (Q1*Q2) to produce the average time spent on each study method for the overall data.

\subsection{Results}

\begin{figure}
  \includegraphics[width=0.95 \columnwidth]{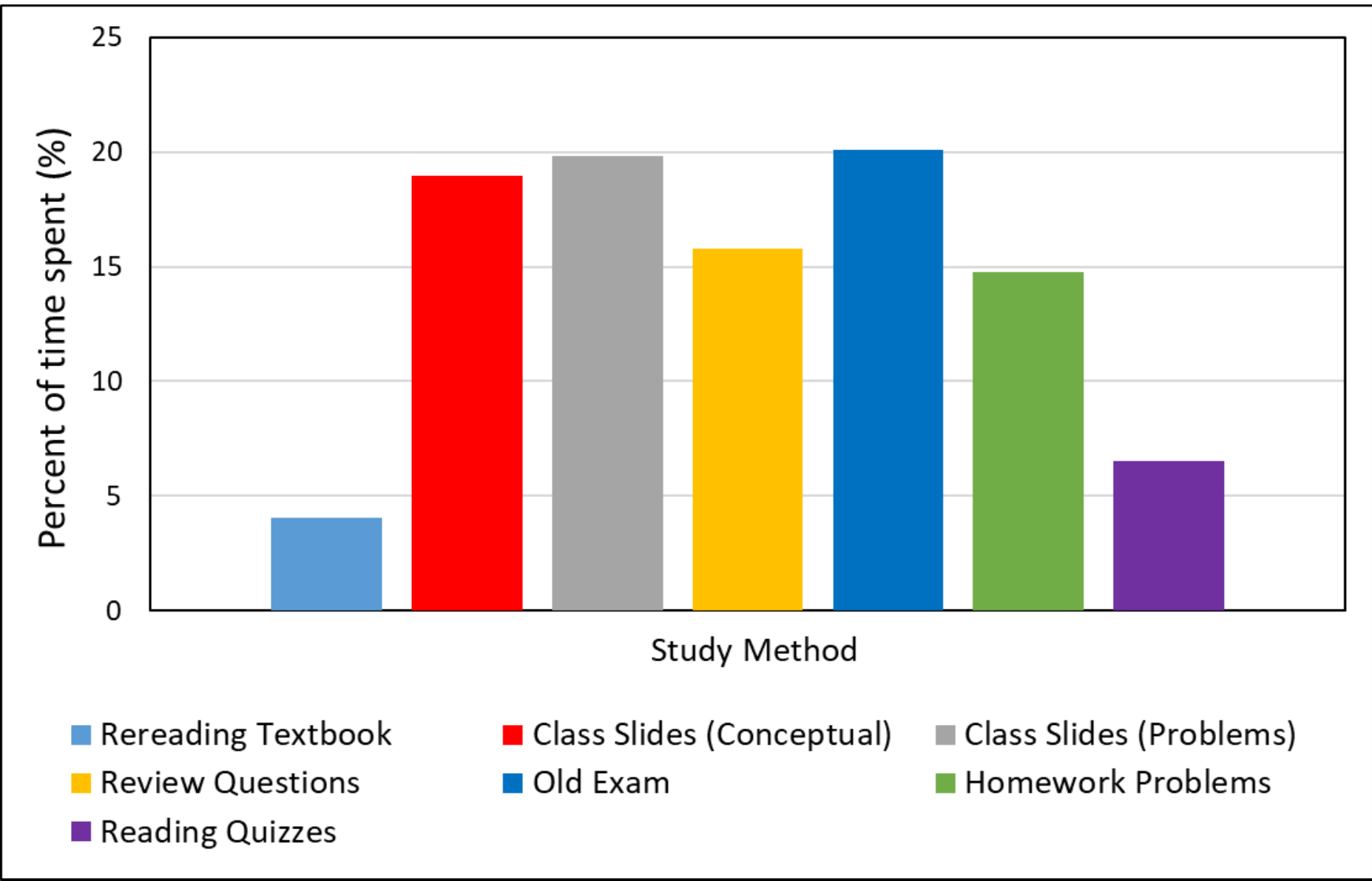}
  \caption{Average percentage of time spent studying. The average percentage of time that students spent on each study method. Out of the 7 study methods, students spent the least percentage of their overall study time on rereading the textbook and reading quizzes. The most popular study methods were class slides (problems), class slides (conceptual), and old exams.} 
  \label{PercTime}
\end{figure}

To determine the overall effectiveness and use of each study method, we compared the different graphs produced. As seen in Figure \ref{PercTime}, on average, students spent a majority percentage of time studying using class slides (problems), old exams, and class slides (conceptual). The average amount of time that students spent rereading the textbook was surprisingly small.

The graph for average percentage of time studying (Figure \ref{PercTime}) was produced with the averaged data. Although we calculated the total time spent by multiplying the total time that each student spent studying by the percentage that each study method was used (Q1*Q2), the graph looked nearly identical to Figure \ref{PercTime}, so we did not include the graph for the total time spent on each study method. Also, when we discuss comparisons, we find that the correlation results are similar for both the percentage of time and the total time spent on each study method.

A lingering question throughout this study is whether, if students know they will be asked about their study methods (which they realize after the first exam in the fall semester), this has an effect on how they prepare for subsequent exams.

\section{Study Method Helpfulness}
\subsection{Data Analysis}

The perceived helpfulness of each study method per exam per semester from Q3 was averaged. This was done in a similar method to how the average total amount of time that each student spent studying for each exam was calculated. Then, the helpfulness of each study method for every semester was averaged to find the overall average perceived helpfulness for each study method. We also took the average of the helpfulness rating for all of the study methods over all of the exams throughout all of the semesters. This process gave us an overall average helpfulness rating.

We used grouping of study methods to compare how students could have studied. Study methods that were more helpful were grouped together, as were study methods that were less helpful. During data collection, only those students who marked that they had used a certain study method in Q2 also recorded the helpfulness of that study method in Q3. Students who did not use a study method did not include a helpfulness rating. This means that if a study method had a low helpfulness rating, that was because the students who used that method had rated it poorly. We are assuming, although we have no evidence, that those who did not use a study method chose to ignore this option because they decided it would not have been helpful, and thus rarely used study methods probably would have had an even lower helpfulness rating if all respondents had been asked to rate the helpfulness of all study methods.

We used three different ways to determine whether a study method was helpful or unhelpful. The first was to look at how students individually rated the helpfulness of study methods. The second was to look at how students rated the helpfulness of study methods each semester on average. The third way was to create groups of universally helpful and unhelpful study methods.

For individual helpfulness (individual), we compare how a student rated each study method for that exam to the overall average helpfulness rating. A study method is considered helpful to that student if they rated the study method as higher than the overall average, and vice versa. In this case, we cannot create separate groups for study methods that are helpful vs. unhelpful because each student had a different interpretation of which study methods were helpful for them.

For average helpfulness (average), if a study method for a specific exam in a specific semester received an average helpfulness that was greater than the overall average, then that study method was determined to be helpful and was placed in the helpfulness group. Similarly, the unhelpful study methods received less than the overall average helpfulness rating. In this case, the helpful and unhelpful study methods were the same for all students.

For helpfulness groups (group), any study method that was more helpful than this overall helpfulness rating after averaging all of the semesters together was labeled helpful, and any method that had a lower-than-average helpfulness was labeled unhelpful.

\subsection{Results}

\begin{figure}
  \includegraphics[width=0.95 \columnwidth]{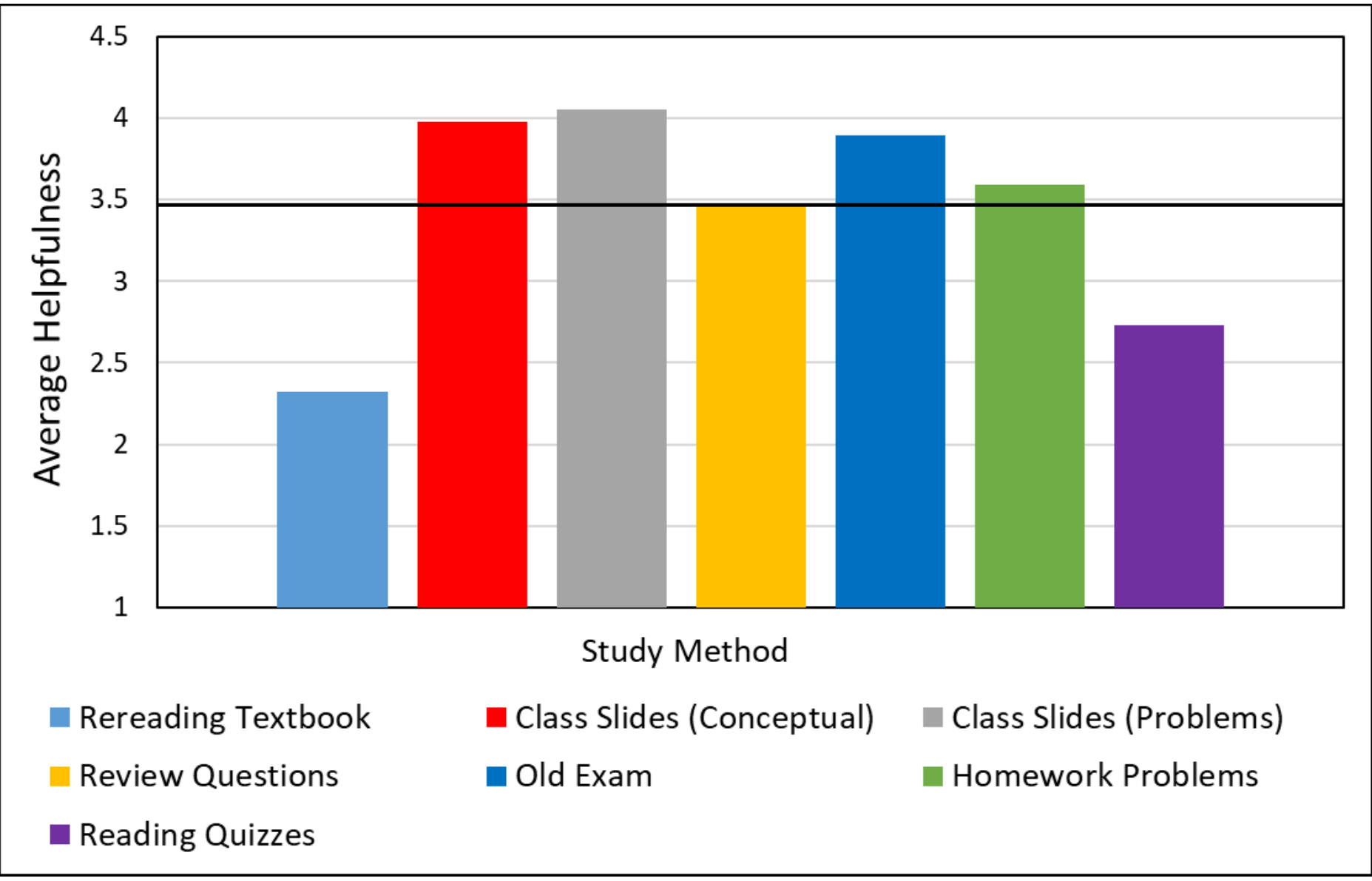}
  \caption{Study Method Helpfulness. The helpfulness rating for each study method averaged over every exam in every semester. The horizontal black line represents the overall helpfulness rating. Of the 7 study methods, four are clearly above the overall helpfulness rating of 3.44 and two are clearly below. This demarcation allows us to designate these study methods into helpful and unhelpful groups. Review questions is at the overall helpfulness rating and is not included into either designation.} 
  \label{Helpful}
\end{figure}

To determine the overall average, we took all of the helpfulness ratings for every study method used by every student for every exam in every semester and averaged all of these to obtain an overall helpfulness rating. The overall helpfulness rating (OHR) was 3.44 out of a total of 5. Any study method that was rated higher than this value was determined to be helpful for that student (individual) or during that exam (average) or overall (group). In order to recognize if a study method was helpful or unhelpful overall, we can look at Figure \ref{Helpful} to determine the overall helpfulness of each study method, which uses the averages from Q3. The horizontal black line represents the overall helpfulness rating. We can see that class slides (both types), the old exams, and homework problems were all higher than the OHR. These study methods are considered helpful to most students and, are placed in the helpful group in this paper. On the other hand, rereading the textbook and reading quizzes had an overall helpfulness that was lower than the OHR, are considered less helpful to most students, and are placed in the unhelpful group.

\begin{figure}
  \includegraphics[width=0.95 \columnwidth]{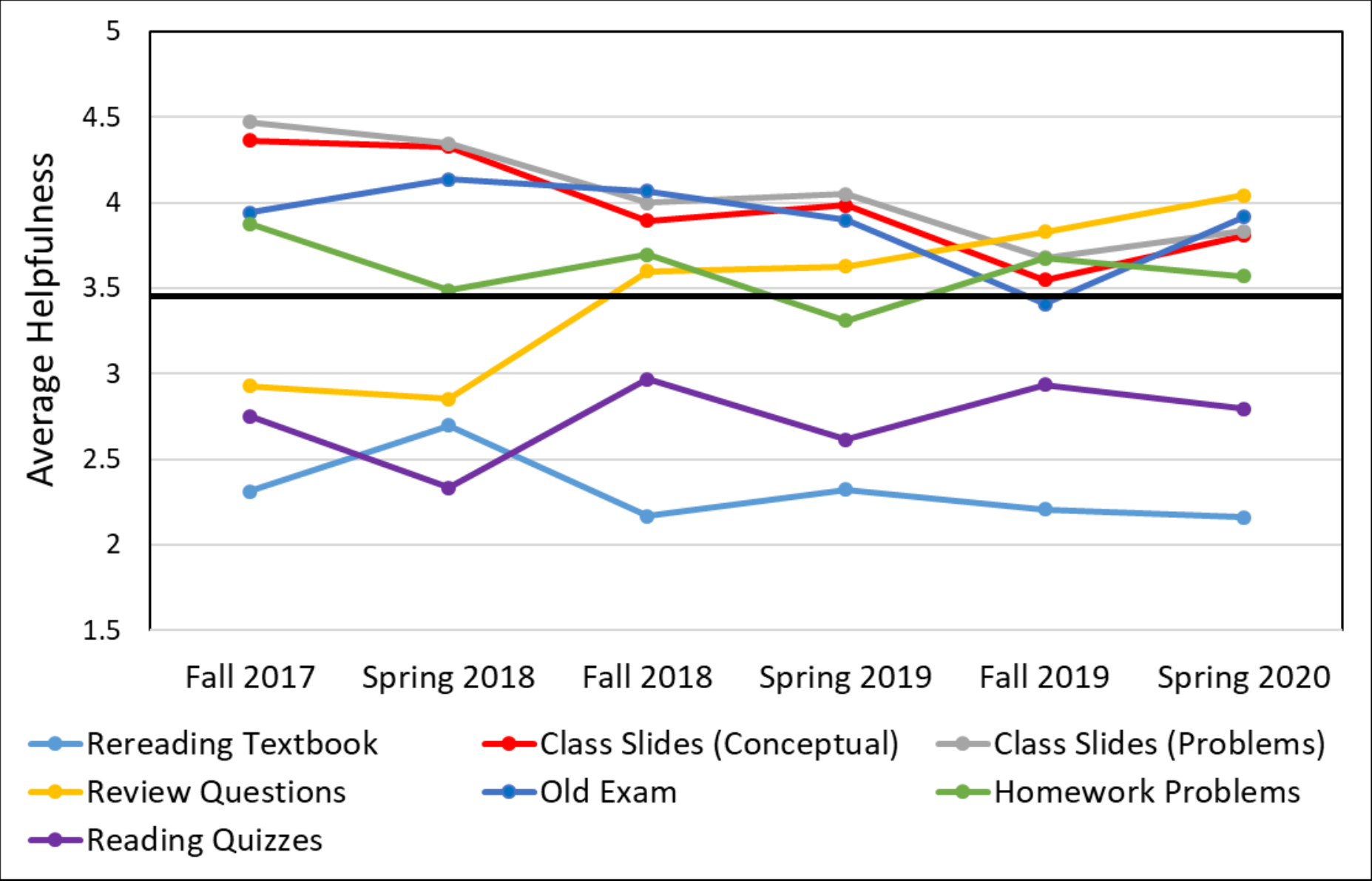}
  \caption{Study Method Helpfulness for semester. Each series indicates how a study method changed throughout the 6 semesters. To determine these values, all of the helpfulness ratings for each student and each exam throughout the semester are averaged together.} 
  \label{HelpfulSem}
\end{figure}

To get an idea of how the study methods’ helpfulness changes throughout each semester, we can look at Figure \ref{HelpfulSem}, in which the data was made into a line graph to help demonstrate the different helpfulness for each study method throughout each semester. Once again, a horizontal black line is provided as a guide to represent the OHR. We can see that for the most part, a study method that starts off as helpful remains that way. The only study method that does not follow this rule is review questions. This could be because many more questions were added to the review questions each year and improvements were made to these questions. Because review questions shifted throughout the semesters, as seen in Figure \ref{HelpfulSem}, it cannot be designated as always helpful or unhelpful.

The Average Helpfulness graphs, both the bar graph (Figure \ref{Helpful}) as well as the line graph (Figure \ref{HelpfulSem}), indicate the perceived helpfulness of each study method. Class slides (conceptual) and class slides (problems) seemed to be deemed the most helpful when preparing for an exam. Both of these study methods received the highest rating of all the different study methods; 4 out of 5. Those that used rereading textbook as a study method did not find that study method to be as helpful to their performance on an exam compared to the other study methods. This observation can be strengthened by Figure \ref{HelpfulSem}, which indicates a downward trend of ratings of helpfulness from the first semester that this survey was administered to its last. Comparing the average time that students spent on each study method as well as the students’ perception of the helpfulness of each study method helped us understand the potential impact of each study method.

Comparing Figure \ref{PercTime} with Figure \ref{Helpful}, we find that students spend more time using study methods that most of the class believed are helpful. Perhaps unsurprisingly, the most helpful study methods are ones that involve questions and problems that are similar to exam questions and problems. In the next section, we compare the effect that relying more on the helpful study methods had on exam grades.

\section{Comparisons}
\subsection{Data Analysis}

To determine how the use of each study method correlated with exam grades, we wrote a program in R that calculated the correlation value and the significance of this correlation. We used the rcorr function in R using the Pearson correlation method. The inputs were the groups of study methods and the exam grades.

The three different helpfulness categories were the individual helpful and unhelpful study methods (individual), the average helpfulness (average), and the grouping of study methods into helpful and unhelpful (group). These groupings allowed us to compare the grades of students who spent more of their study time using the 4 helpful study methods and those who relied more on the two unhelpful study methods. As a reminder, the groupings were as follows:

\begin{itemize}
\item Helpful Study Methods
   \begin{itemize}

   \item Class Slides (Conceptual)
   \item Class Slides (Problems)
   \item Old Exam
   \item Homework Problems
   \end{itemize}

\item Unhelpful Study Methods
   \begin{itemize}

   \item Rereading Textbook
   \item Reading Quizzes
\end{itemize}
\end{itemize}

The study method labeled as helpful for one student (individual) or one exam (average) may not be labeled as helpful every time. Figure \ref{HelpfulSem} offers an overview of the helpfulness of each study method in a specific semester. For the groupings, the same study methods are labeled helpful or unhelpful for all exams.

We broke the exam grades into three categories. First, we used the overall exam averages. Because the exams were split between mathematical problems and conceptual questions, we further broke the exam grades out into problems (2nd category) and conceptual (3rd category).

All reported correlations have a significance of p<0.05. If a value from the correlation table had a p-value of greater than or equal to 0.05, we did not include that value in our results. If a value included a significance of more than one star, then 2 stars represents p<0.01 and 3 stars represents p<0.001. Since there were 21 exams during the 6 semesters in which data was collected, we will be discussing the correlations that show up in many of the exams.

\subsection{Results}

To determine if there was a correlation, we did not look at only one exam or only one semester, we looked at every exam from every semester. For the individual helpfulness rating, significant correlations showed up in 13 of the 21 exams. For the average helpfulness, significant correlations showed up in 7 of the exams. For the groupings, significant correlations showed up in 11 of the exams. We concluded from this that there is strong evidence that individual helpfulness correlates with exam grades, there is evidence that there is a correlation between helpfulness groups and exam grades, and there is weaker evidence that there is a correlation between average helpfulness and exam grades.

\begin{figure}
  \includegraphics[width=0.95 \columnwidth]{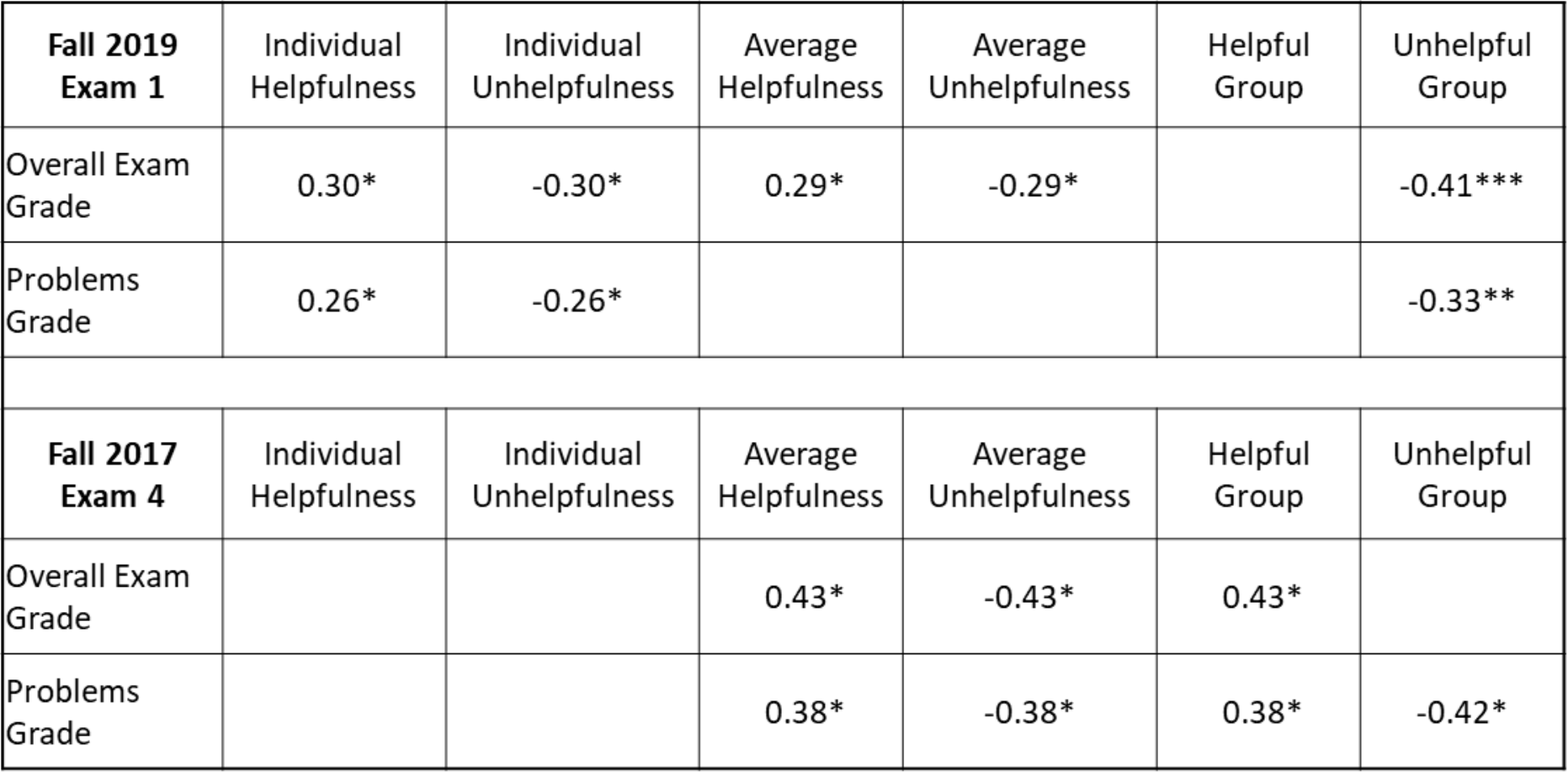}
  \caption{Correlation table example from the first exam in the fall of 2019 and fourth exam in the fall of 2017 showing that there were significant correlations between individual helpfulness and average helpfulness ratings and exam grades. Stars represent the level of significance and blank spaces represent correlations that were not significant. Exam grades are broken down further into the grades for the mathematical problems and the conceptual questions.} 
  \label{Corr}
\end{figure}

An example of correlation data that shows this relationship is shown in Figure \ref{Corr}. These tables represent only 2 of the 21 exams, but they are indicative of the strong correlations that we find between using helpful study methods and exam grades. We see in Exam 1 that there is a significant positive correlation for individual helpfulness (individual) across all overall exam grades and exam grades from problems. These correlations indicate that when a student spent a greater percentage of their time on study methods that they labeled as more helpful than the OHR, the student’s grades were higher. Exam 1 also shows a significant negative correlation between spending more time on individually unhelpful study methods and exam grades, which means students’ exam grades were lower after using less helpful study methods. We see in Exam 4 that there are significant correlations for average helpfulness (average) and helpfulness groupings (group) for both overall exam grades and the grades for mathematical problems. We did not include conceptual grades in these examples because there were fewer correlations.

exams by spending more time on study methods that they rated as more helpful and less time on study methods that they rated as unhelpful (individual) had higher grades on their exams. Students who spent more time on study methods that they labeled as unhelpful and less time on study methods that they labeled as helpful had lower exam grades.

A similar result is found for students who studied for exams by spending more time on helpful study methods (group). An example of these correlations is shown in Exam 4 of Figure \ref{Corr}. Exam 4 is only a single example, but it is indicative of the evidence that we find. We see that there is a significant positive correlation between using study methods that are labeled on average as more helpful and the overall exam grades and the grades from problems. We did not see as many correlations between using helpful study methods and the grades on the conceptual parts of the exam, so this was not included in the table.

Although weaker, there is some evidence that students who spend more time on study methods labeled as helpful for that exam (average) earned higher exam grades. Examples of these correlations are shown in Exams 1 and 4 of the table in Figure \ref{Corr}. Because this is weaker, it is possible that study methods may be rated differently for each exam, despite their overall helpfulness not changing much.

By combining all three methods (individual, average, and group), we find significant correlations between studying using helpful study methods and exam grades on all but 6 of the 21 exams. These exams are spread over fall and spring and over each of the exams, so there is not anything specific that we can point to explaining why some exams did not show any significant correlations.

\section{DISCUSSION}

We find a positive correlation between the time spent studying using study methods that an individual student labeled as helpful (individual) and higher exam grades. We also find a positive correlation between the time spent using helpful study methods (group) and higher exam grades. When students study for exams using more helpful study methods, they end up doing better on their exams. Unfortunately, the definition of what makes a study method helpful for each student is difficult to determine. The helpful group suggests that we could provide future students with information about the helpful and unhelpful study methods. We could imagine that providing a suggestion of what material will be most helpful in studying for exams might help students to improve their exam scores, especially those who may not be as good at differentiating which study methods are helpful. On the other hand, the evidence from individual helpfulness suggests that students who study what they determine to be helpful have higher grades, and vice versa. For this reason, providing the average helpfulness ratings could actually hinder a student who finds something helpful that others did not.

After finding evidence that students who study using more helpful study methods do better on their exams, we wonder about the causal relationship. Is this due to the fact that some students are better at choosing helpful study methods? Are they better at recognizing what was most helpful for them or did the study methods themselves help the students? Answering these question is the key to truly understanding what these results mean and what to do with them. For this reason, further research will be needed. One possibility would be to provide information about the helpful group of study method (group) to one group of students and not to a control group. We could track whether having this information had an effect on their choice of study methods and their exam grades. While we were hoping to come away from this project with a definitive list of helpful study methods to disseminate to current and future students, we recognize that more research is needed.

\end{document}